\documentclass[12pt]{article}

\pdfoutput=1

\usepackage{cite}

\usepackage{amssymb,latexsym}

\usepackage{epstopdf}

\usepackage{graphics,epsfig}

\usepackage{color}

\usepackage{amsmath,amsfonts}

\usepackage{mathrsfs}

\DeclareMathAlphabet{\mathpzc}{OT1}{pzc}{m}{it}

\usepackage{tikz}
\usetikzlibrary{arrows,shapes}
\usetikzlibrary{trees}
\usetikzlibrary{matrix,arrows} 				
\usetikzlibrary{positioning}				
\usetikzlibrary{calc,through}				
\usetikzlibrary{decorations.pathreplacing}  
\usepackage{pgffor}							

\usetikzlibrary{decorations.pathmorphing}	
\usetikzlibrary{decorations.markings}
\tikzset{
    vector/.style={decorate, decoration={snake}, draw},
	provector/.style={decorate, decoration={snake,amplitude=2.5pt}, draw},
	antivector/.style={decorate, decoration={snake,amplitude=-2.5pt}, draw},
    fermion/.style={draw=black, postaction={decorate},
        decoration={markings,mark=at position .55 with {\arrow[draw=black]{>}}}},
    fermionbar/.style={draw=black, postaction={decorate},
        decoration={markings,mark=at position .55 with {\arrow[draw=black]{<}}}},
    fermionnoarrow/.style={draw=black},
    gluon/.style={decorate, draw=black,
        decoration={coil,amplitude=4pt, segment length=5pt}},
    scalar/.style={dashed,draw=black, postaction={decorate},
        decoration={markings,mark=at position .55 with {\arrow[draw=black]{>}}}},
    scalarbar/.style={dashed,draw=black, postaction={decorate},
        decoration={markings,mark=at position .55 with {\arrow[draw=black]{<}}}},
    scalarnoarrow/.style={dashed,draw=black},
    electron/.style={draw=black, postaction={decorate},
        decoration={markings,mark=at position .55 with {\arrow[draw=black]{>}}}},
	bigvector/.style={decorate, decoration={snake,amplitude=4pt}, draw},
}

\tikzstyle{block} = [draw, rectangle, 
    minimum height=3em, minimum width=6em]

\let\a=\alpha \let\b=\beta \let\g=\gamma \let\d=\delta \let\e=\epsilon
\let\z=\zeta  \let\th=\theta  \let\k=\kappa
\let\l=\lambda \let\m=\mu \let\n=\nu \let\x=\xi \let\p=\pi 
\let\s=\sigma   \let\f=\phi  
 
        \let\Th=\Theta 
\let\X=\Xi  \let\S=\Sigma  \let\Y=\Psi
 
\let\la=\label  
  
\def\nn{\nonumber} \def\bd{\begin{document}} \def\ed{\end{document}}
\def\ds{\documentstyle} \let\fr=\frac \let\bl=\bigl \let\br=\bigr
\let\Br=\Bigr \let\Bl=\Bigl
\let\bm=\bibitem
\let\na=\nabla
\def\tU{{\widetilde U}}
\let\pa=\partial \let\ov=\overline
\def\ie{{\it i.e.\ }}
\newcommand{\be}{\begin{equation}}
\newcommand{\ee}{\end{equation}}
\def\ba{\begin{array}}
\def\ea{\end{array}}
\def\ft#1#2{{\textstyle{{\scriptstyle #1}\over {\scriptstyle #2}}}}
\def\fft#1#2{{#1 \over #2}}
\def\F#1#2{{ F_{#1}^{(#2)} }}
\def\cF#1#2{{ {\cal F}_{#1}^{(#2)} }}

\def\R{{\bf R}}
\def\sst#1{{\scriptscriptstyle #1}}
\def\oneone{\rlap 1\mkern4mu{\rm l}}
\def\e7{E_{7(+7)}}
\def\td{\tilde}
\def\wtd{\widetilde}
\def\im{{\rm i}}
\def\bog{Bogomol'nyi\ }
\newcommand{\ho}[1]{$\, ^{#1}$}
\newcommand{\hoch}[1]{$\, ^{#1}$}
\newcommand{\bea}{\begin{eqnarray}}
\newcommand{\eea}{\end{eqnarray}}
\newcommand{\ra}{\rightarrow}
\newcommand{\lra}{\longrightarrow}
\newcommand{\Lra}{\Leftrightarrow}
\newcommand{\ap}{\alpha^\prime}
\newcommand{\bp}{\tilde \beta^\prime}
\newcommand{\cB}{{\cal B}}
\newcommand{\cO}{{\cal O}}
\newcommand{\vecx}{\vec{x}}
\newcommand{\vecy}{\vec{y}}
\newcommand{\vecp}{\vec{p}}
\newcommand{\vecq}{\vec{q}}
\newcommand{\tr}{{\rm tr} }
\newcommand{\Tr}{{\rm Tr} }
\newcommand{\NP}{Nucl. Phys. }

\newcommand{\cL}{{\cal L}}
\newcommand{\cA}{{\cal A}}
\newcommand{\cT}{{\cal T}}
\newcommand{\cR}{{\cal R}}
\newcommand{\cD}{{\cal D}}
\newcommand{\cH}{{\cal H}}

\def\Cb{\bar{C}}

\def\sst#1{{\scriptscriptstyle #1}}
\def\0{{\sst{(0)}}}
\def\1{{\sst{(1)}}}
\def\2{{\sst{(2)}}}
\def\3{{\sst{(3)}}}
\def\4{{\sst{(4)}}}
\def\5{{\sst{(5)}}}
\def\6{{\sst{(6)}}}
\def\7{{\sst{(7)}}}
\def\8{{\sst{(8)}}}
\def\9{{\sst{(9)}}}
\def\p{{\sst{(p)}}}
\def\q{{\sst{(q)}}}
\def\ve{\varepsilon}
\def\vf{\varphi}
\def\F{\Phi}
\def\wg{\wedge}

\def\thb{\bar{\theta}}
\def\Thb{\bar{\Theta}}
\def\barp{\bar{p}}
\def\barq{\bar{q}}
\def\barc{\bar{c}}
\def\bard{\bar{d}}
\def\e{\epsilon}

\def \bi{\bibitem}
\def \la {\label}

\def \l {\lambda}
\def\foot{\footnote}
\def \tl  {{\tilde \l}}
\def \sql {{\sqrt \l}}
\def \adss {$AdS_5 \times S^5$\ }
\newcommand{\rf}[1]{(\ref{#1})}
\def \ov {\over}

\def\th{\theta}
\def\Th{\Theta}
\def\vth{\vartheta}
\def\btheta{{\bar\theta}}
\def\ttheta{{{\tilde\theta}}}
\def\bttheta{{{\bar\ttheta}}}
\def\vth{\vartheta}

\def\ra{\rightarrow}
\def\N{\nabla}
\def\F{{\cal F}}
\def\uM{\underline{M}}
\def\uA{\underline{A}}
\def\uN{\underline{N}}
\def\uP{\underline{P}}
\def\ua{\underline{a}}
\def\ub{\underline{b}}
\def\uc{\underline{c}}
\def\ud{\underline{d}}
\def\ue{\underline{e}}
\def\uf{\underline{f}}
\def\ui{\underline{i}}
\def\uj{\underline{j}}
\def\uk{\underline{k}}
\def\ul{\underline{l}}
\def\ual{\underline{\alpha}}
\def\ube{\underline{\beta}}
\def\um{\underline{m}}
\def\un{\underline{n}}
\def\up{\underline{p}}
\def\uq{\underline{q}}
\def\ur{\underline{r}}
\def\us{\underline{s}}
\def\umu{\underline{\mu}}
\def\unu{\underline{\nu}}
\def\ula{\underline{\l}}
\def\uka{\underline{\k}}
\def\usi{\underline{\s}}
\def\urh{\underline{\r}}
\def\cc{\circ}
\def\eqv{\equiv}

\def\ni{\noindent}

\def\Ep{E^{{}^{(+)}}}
\def\Em{E^{{}^{(-)}}}

\def\Mp{M^{{}^{(+)}}}
\def\Mm{M^{{}^{(-)}}}

\def \ha{{1\ov 2}}

\def\r{\rho}

\def\Y{{\rm Y}}
\def\X{{\rm X}}
\def\tY{\tilde{\rm Y}}
\def\tX{\tilde{\rm X}}
\def\dY{\dot{\rm Y}}
\def\dX{\dot{\rm X}}

\def \J {\mathcal{J}}
\def \del {\partial}

\def\dF{\dot{F}}
\def\dG{\dot{G}}
\def\df{\dot{f}}
\def \E {{\cal E}}
\def \S {{\cal S}}
\def \J {{\cal J}}

\def\ms{\mathcal{S}}
\def\mj{\mathcal{J}}
\def\soj{\fr{\ms}{\mj}}
\def \R {{\bf R}}
\def \om {\omega}
\def \bE {\bar E}
\def \x {{\cal X}}

\def \bi{\bibitem}
\def \la {\label}

\def \l {\lambda}
\def\foot{\footnote}
\def \tl  {{\tilde \l}}
\def \sql {{\sqrt \l}}
\def \adss {$AdS_5 \times S^5$\ }
\def \ov {\over}

\def \varpi {{\rm w}}

\def\thb{\bar{\theta}}
\def\Thb{\bar{\Theta}}
\def\mb{\bar{\m}}
\def\ab{\bar{\a}}
\def\zb{\bar{z}}
\def\psib{\bar{\psi}}
\def\barp{\bar{p}}
\def\barq{\bar{q}}
\def\barc{\bar{c}}
\def\bard{\bar{d}}
\def\e{\epsilon}
\def\wb{\bar{w}}
\def\lb{\bar{\l}}
\def\Jb{\bar{J}}
\def\Nb{\bar{N}}
\def\Zb{\bar{Z}}
\def\pab{\bar{\pa}}

\def\At{\tilde{A}}
\def\Bt{\tilde{B}}
\def\Ct{\tilde{C}}
\def\Dt{\tilde{D}}
\def\Et{\tilde{E}}
\def\Ft{\tilde{F}}
\def\Gt{\tilde{G}}
\def\Ht{\tilde{H}}
\def\Kt{\tilde{K}}
\def\Mt{\tilde{M}}
\def\Nt{\tilde{N}}
\def\Rt{\tilde{R}}
\def\at{\tilde{a}}
\def\bt{\tilde{b}}
\def\ct{\tilde{c}}
\def\dt{\tilde{d}}
\def\et{\tilde{e}}
\def\ft{\tilde{f}}
\def \ztt{\tilde{\z}}
\def\htil{\tilde{h}}
\def\gt{\tilde{g}}
\def\nt{\tilde{n}}
\def\mut{\tilde{\mu}}
\def\nut{\tilde{\nu}}
\def\pht{\tilde{\f}}
\def\vft{\tilde{\vf}}

\def\rht{\tilde{\rho}}

\def\asth{\hat{*}}
\def\phh{\hat{\phi}}

\def\bA{{\bf A}}

\def\ola{\overleftarrow}
\def\ora{\overrightarrow}
\def\alt{\tilde{\a}}

\def\eh{\hat{e}}
\def\eph{\hat{\e}}
\def\ph{\hat{p}}
\def\alh{\hat{\a}}
\def\beh{\hat{\b}}
\def\gah{\hat{\g}}
\def\Fh{\hat{F}}
\def\muh{\hat{\m}}
\def\nuh{\hat{\n}}
\def\thh{\hat{\th}}
\def\rhh{\hat{\r}}
\def\dh{\hat{d}}
\def\ih{\hat{i}}
\def\jh{\hat{j}}
\def\hh{\hat{h}}
\def\nh{\hat{n}}
\def\gh{\hat{g}}
\def\kh{\hat{k}}
\def\deh{\hat{\d}}
\def\wh{\hat{w}}
\def\lah{\hat{\l}}
\def\Ah{\hat{A}}
\def\Kh{\hat{K}}
\def\Nh{\hat{N}}
\def\Rh{\hat{R}}
\def\Ch{\hat{C}}
\def\Omh{\hat{\Omega}}

\def\xh{\hat{x}}

\def\ps{\rlap{\, /}\;\,p }
\def\ks{\rlap{\, /}\;\,k }

\def\gym{g_{YM}}

\def\adot{\dot{a}}
\def\bdot{\dot{b}}
\def\bpa{\bar{\pa}}

\def\pr{\prime}
\def\ssk{\medskip}
\def\clb{\color{blue}}
\def\clr{\color{red}}
\def\clg{\color{green}}

\def\bfA{{\bf A}}
\def\bfB{{\bf B}}
\def\bfK{{\bf K}}
\def\bfU{{\bf U}}
\def\bfX{{\bf X}}
\def\bfY{{\bf Y}}
\def\bfZ{{\bf Z}}
\def\bfg{{\bf g}}
\def\bfn{{\bf n}}

\begin{document}

\overfullrule=0pt
\parskip=2pt
\parindent=12pt
\headheight=0in \headsep=0in \topmargin=0in
\oddsidemargin=0in

\vspace{ -3cm}
\thispagestyle{empty}

 \vspace{0.1cm}

\setcounter{equation}{0}
\setcounter{footnote}{0}
\setcounter{section}{0}

\begin{center}

{\Large\bf Foliation-based quantization and black hole information}

\vskip 0.8cm

 \vspace{.5cm}

\vspace{0.5cm}
I. Y. Park
\\

\vspace{0.3cm}

\vspace{0.3cm}
{\it Department of Applied Mathematics,
Philander Smith College 
                               \\
Little Rock, AR 72223, USA \\
inyongpark05@gmail.com
}

\end{center}

 \vspace{0.1cm}

\begin{abstract}

We extend the foliation-based quantization scheme of \cite{Park:2014tia} to arbitrary asymptotically flat backgrounds including time- and position- dependent ones.
One of the ingredients to accomplish the extension is imposition of a Neumann-type boundary condition. The quantization procedure, especially the gauge-fixing-induced reduction, provides a new insight into the black hole information paradox. The hypersurface degrees of freedom in the asymptotic region - whose dynamics should be responsible for part of the `hair' - and transitions among various excitations play a central role in the global formulation of the information and proposed solution of the information paradox. In retrospect, the quantization scheme reveals the origin of the difficulty of the information problem: the problem's ties with the quantization of gravity and subtle boundary dynamics as well as the multilayered techniques required for its setup and study.  We also comment on the implications of  the asymptotic symmetries for the present quantization framework.

\end{abstract}
\newpage

\section{Introduction}

The quantization of gravity (see \cite{DeWitt:1975ys,'tHooft:1974bx,Brill:1970df,Stelle:1976gc,Antoniadis:1986tu,Weinberg3,Reuter:1996cp,Odintsov:1991yx,Barvinsky:1993zg,Carlip:2001wq,Ambjorn:2012jv,Woodard:2014jba,Donoghue:2015hwa} for various approaches) should either hold a direct key or provide a route to a solution of some of the longstanding problems of theoretical physics, such as the black hole information paradox \cite{Hawking:1976ra,Page:1993up,Mathur:2009hf,Hooft:2016vug,Polchinski:2016hrw}. In the present study we extend the foliation-based quantization scheme put forth in \cite{Park:2014tia,Park:2015ybl,Park:2014noa,Park:2015xoa,Park:2016zgt}\footnote{We have recently become aware, to our pleasant surprise, that these works are quite in spirit of \cite{Brill:1970df}.} to more general backgrounds, such as those that are time- and position- dependent. 
Up to now the method has been applied to relatively simple backgrounds including a flat or Schwarzschild or de Sitter background. Although a substantial number of interesting and important backgrounds have been covered, it has not been clear whether or not the method is applicable to, say, a time- and position- dependent one. Here we improve the status of this matter, and make the applicable range of the scheme more precise: we show that the method is applicable (but not limited) to arbitrary backgrounds with asymptotical flatness.

One of the driving themes of the recent developments in theoretical physics, in particular, string theory, has been holography. In the present study we take another detailed look at the roles played by the boundary conditions\footnote{There exists a long list of the studies of the boundary degrees of freedom in gravity theories (see, e.g., \cite{Benguria:1976in,Witten:1988hf,Balachandran:1991dw,Smolin:1995vq}). Recent studies of boundary conditions can be found, e.g., in \cite{Parattu:2016trq,Krishnan:2016mcj,Lehner:2016vdi}.} and boundary's degrees of freedom. We {\em explicitly} demonstrate how the boundary develops its own dynamics from the bulk dynamics. Since the `dual' boundary degrees of freedom arises from a Kaluza-Klein type procedure, their appearance can be directly seen, which offers great conceptual and technical advantages over the common practices in AdS/CFT-type dualities.

We then apply the quantization scheme to the black hole information (BHI) and cogitate its consequence. In particular, we corroborate and refine the picture presented earlier in \cite{Park:2013rm} in which a certain pattern of black hole information release was envisaged.
One of the compelling motivations of the present study of BHI is the following observation: the Dirichlet boundary condition widely used in gravity theories seems at odds with the the dual boundary {\em dynamism}.  
The quantization scheme of \cite{Park:2014tia,Park:2016zgt} poses a similarly pressing question: the physical states of a gravity theory should be associated with the hypersurface in the asymptotic region. Wouldn't it imply an active and dynamic role of the hypersurface? One may go further and ask a question at a more fundamental level: shouldn't the physical degrees of freedom at the hypersurface play an essential role in formulating the meaning of the information and its (re)distributions (or ``loss") \cite{Baierlein:1962zz}?  The quantization scheme suggests a natural solution to the non-dynamism problem by adopting a Neumann-type boundary condition \cite{Park:2016fxc}. Below we also address the other questions either by directly analyzing them or outlining the steps executable with a reasonable amount of efforts.

As we will describe in detail in the main body, the BHI is a highly multilayered problem and requires several less familiar techniques, some of which have become available only quite recently. Quantization of gravity has been the most severe obstruction to setting up the problem at a formal and fully second-quantized level. The quantization procedure of \cite{Park:2014tia,Park:2015ybl,Park:2016zgt} that we will follow includes the steps of identifying the physical degrees of freedom. The association of the physical states with the hypersurface at the asymptotic region suggests that the in- and out- states constructed out of the hypersurface degrees of freedom\footnote{From the bulk point of view, these states must be associated with the past and future null infinities.} will represent the bulk propagating states projected onto the holographic screen, the hypersurface.

As a matter of fact, a clear understanding of the boundary degrees of freedom greatly adds towards the insight  of a global picture of the information problem.  
Let us consider the case of a Schwarzschild black hole, to be specific. According to the analysis carried out in \cite{Park:2015xoa,Park:2014tia}, the physical states of the bulk admits an holographic description through the dynamics of a certain hypersurface. At the moment, the precise specification of the hypersurface is irrelevant and will be postponed for the sake of a quick discussion; we will have a detailed discussion in the main body.     
Now consider two different slicings such as the Schwarzschild coordinates and Kruskal coordinates. 
One would impose a Dirichlet boundary condition for the boundary hypersurface for each slicing.
Since the two Dirichlet boundary conditions will not be connected by the coordinate transformation and the hypersurface degrees of freedom are quite essential to the system, this situation hints at the need to enlarge the theory's Hilbert space, a topic to which we now turn.  

Recently there has been a proposal  \cite{Freidel:2016bxd,Donnelly:2016auv} that all of those different boundary conditions must be included in an enlarged Hilbert space, a standpoint that we adopt and generalize in the present work. Furthering the proposal, we will argue that in general, selecting a boundary condition should be viewed as {\em part of making an ansatz} for the solution or a class of the solutions of the system, all of which belong to the enlarged Hilbert space.
We will compare this view with that of the studies of the asymptotic symmetry, especially the BMS group \cite{Bondi:1962px,Sachs:1962wk}, recently advanced with renewed interest.

In the analysis below we employ three different second quantized descriptions. The standard Lagrangian formalism is optimal for handling the  scattering process of the Fock states. For determination of the physical states we use the ADM Hamiltonian and Lagrangian formalisms \cite{Arnowitt:1962hi}. In addition, the second-quantized Schrodinger description (SQSD) offers certain conceptual advantages in dealing with the vacuum transition and other related aspects.     
In setting up and studying the BHI in SQSD, one's focus lies in the analysis of the transitions between different quantum states of the `combined' system of the bulk and boundary.

\vspace{.2in}
In section 2 we start with a guiding review of the foliation-based quantization scheme.
Then we extend the method to a more general class of backgrounds with asymptotic flatness. The quantization procedure naturally brings up the issues of the boundary conditions and boundary dynamics. 
With the extension of the quantization scheme completed in section 2, we consider in section 3 a Schwarzschild background to illustrate how to set up the BHI problem. 
While we are at it, we solve a puzzle encountered along the way regarding the disintegration of a black hole: how could the mass of the black hole {\em change} through the Hawking radiation since it is supposed to be a conserved quantity even at the quantum level? To our view, the answer to this question has not been explicated in the literature. We find the solution in the Neumann boundary condition.
 The results of the present paper quantitatively confirms the picture presented in \cite{Park:2013rm} in which the role of jets, hard or soft, was emphasized.
The central feature of our approach to the BHI problem is that there exists a hierarchy in the excitations of the system. As we will see, things can be placed in a clearer perspective in the Schrodinger-picture second quantization description (SQSD).
We explore as analytically as possible how a Schwarzschild black hole disintegrates with some details on the disintegration process. When we do this, an interesting picture emerges: the vacuum bubble diagrams must be responsible for the disintegration. It should be the information-neutral `radiation' emerging from the vacuum transition that must be the Hawking radiation. Most of the information will be stored in the dynamics in the boundary and horizon vicinity. We conclude in section 4 with summary and future directions.

\section{Quantization in more general backgrounds}


According to a common lore, renormalizability is a local property and one can consider just a constant background to establish it. However, the scheme of quantization of \cite{Park:2014tia,Park:2016zgt} has been applied to the backgrounds with certain simple forms of the coordinate-dependence of the metric. Because the method cannot directly applied, without additional consideration, to a time- and position dependent background as we will review below, it is not clear whether one could conclude the renormalizability in arbitrary (or a large enough class of) backgrounds from the flat case study. 
As we will discuss below, it should be possible to draw such a conclusion if one restricts to an asymptotically flat background. Presumably the analysis can also be applied to a background such as asymptotically dS backgrounds. More generally it should be possible to apply to a background that has a vanishing second fundamental form at the boundary.

To review the gauge-fixing induced reduction of the physical states in the simplest setup, we consider the Einstein-Hilbert system.
One will see why the reduction procedure can be easily applied to certain relatively simple backgrounds whereas it is necessary to take additional consideration for more complicated ones. In general what hampers the renormalizability is appearance of the Riemann curvature tensor (as opposed to 
Ricci scalar or Ricci tensor that can be absorbed by a metric field redefinition). As in the previous works, the gauge-fixing of the lapse and shift vector is a critical initial step toward the renormalizability. Examining the 3+1 split form of the Riemann curvature tensor provides a hint at how to control the proliferation of the second fundamental form. This also makes it clearer that the condition required for renormalizability is the vanishing second fundamental form at the asymptotic region - which should be guaranteed by the asymptotic flatness.

Lastly, we contemplate various implications of the first two subsections.
We consider the Schwarzschild black hole to illustrate the issues surrounding the boundary conditions and dynamics, thereby setting the stage for section 3. By splitting out the radial coordinate $r$ as the evolution coordinate, we note that what the lapse function and shift vector gauge-fixing basically archieves is to reduce the physical states to the holographic screen, namely the hypersurface at $r=\infty$. Since the radial coordinate plays the role of `time,' the procedure naturally suggests to impose a Dirichlet boundary condition {\em along} $r$, `$r$-Dirichlet' boundary condition, which is nothing but the {\em Neumann} boundary condition. The potential relevance of Neumann boundary condition has been anticipated in \cite{Park:2016fxc}. (Incidentally, the asymptotic vanishing of the second fundamental form is consistent with Neumann boundary condition.)\footnote{Interestingly, a Neumann boundary condition has appeared in the analysis in \cite{Freidel:2016bxd} from a different context.} In particular, the Neumann boundary condition, unlike the Dirichlet boundary condition, allows a nontrivial dynamics on the hypersurface, which we will see in more detail.\footnote{Therefore it appears that there are at least three reasons to consider the Neumann boundary condition: boundary dynamism, black hole disintegration, and renormalizability.}

\subsection{review of reduction: flat background example}

From the pre-AdS/CFT days, there have been many indications, direct and indirect, that a gravity theory admits a description by holographically reduced set of degrees of freedom \cite{Benguria:1976in,Witten:1988hf,Balachandran:1991dw,Smolin:1995vq}. The works of  \cite{Park:2014tia,Park:2015ybl,Park:2014noa,Park:2015xoa,Park:2016zgt} explicitly showed by employing the ADM formalism that the physical Fock states around the given background are contained in a hypersurface in the asymptotic region, which in turn has led to renormalizability of those states. Here we present a slightly more streamlined review of the quantization procedure and get ready for its extension in the next subsection.

Considering the (3+1) splitting:
\bea
x^\m\equiv (y^m,x^3), \quad  \m=0,..,3,\; m=0,1,2
\eea
it is well known that the Einstein-Hilbert action
\bea
S_{EH}=\int d^4x \sqrt{-g}\;R  \la{eha}
\eea
can be cast into the ADM form\footnote{The last term, $2\nabla_\a(n^\b \nabla_\b n^\a-n^\a \nabla_\b n^\b)$, is the surface term and will be set aside. For a Schwarzschild black hole, a detailed analysis of this surface term has been carried out in \cite{Park:2015xoa}.} \cite{Poisson}
\bea
S = \int d^4 x\;n\sqrt{-h} \Big[\cR+K^2-K_{mn}K^{mn}+2\nabla_\a(n^\b \nabla_\b n^\a-n^\a \nabla_\b n^\b)\Big]
\eea
where $n$ and $N_m$ denote the lapse function and shift vector respectively; $n^\a$ denotes the unit normal to the boundary.
The second fundamental form $K_{mn}$ is given by 
\be
K_{mn}=\fr1{2n}\left(\mathscr{L}_{{3}} h_{mn}-{\nabla}_m N_{n}
         -{\nabla}_n N_{ m} \right),\qquad K\equiv h^{mn}K_{mn}.
\la{K4defqq}
\ee
$\mathscr{L}_{{3}}$ denotes the Lie derivative along the vector field $\pa_{x^3}$ and $\N_m$ is the 3D covariant derivative. As we will see below, splitting out a spatial direction will lead to a very crucial implication for setting up the scattering process.  
Before analyzing the gauge-fixing, let us examine the bulk part of the ``Hamiltonian" of $x^3$-evolution:
\bea
\!\!\!\!\!\!  H=\int d^3y\left[ n( -h)^{-1/2}({ -}\pi^{mn}\pi_{mn} { +} \fr12 \pi^2)
-n (-h)^{1/2}R^\3-2N_m (-h)^{1/2}\N_n[(-h)^{-1/2}\pi^{mn}]  \right]
\nn\\
\eea
where $\pi^{mn}$ denotes the momentum field,
\bea
\pi^{mn}=\sqrt{{-}h}\;({ -}K^{mn} { +}K h^{mn})
\eea
It is possible to  express the Hamiltonian in terms of the lapse and shift constraints;
omitting the surface terms, one gets
\bea
{\mathscr H}  
  &=& \sqrt{-h}\,\Big[-n {\cal C}-{2}N^{m}{\cal C}_{m} \Big]
\eea
where
\bea
{\cal C}_\equiv \cR-K^2+K_{mn}K^{mn} \quad,\quad
{\cal C}_{m} \equiv \N^n(-K_{mn}+K h_{mn})
\eea
When the second fundamental form is expressed in terms of the momentum field $\pi^{mn}$, the requirements, ${\cal C}=0$ and ${\cal C}_{m}=0$, are called the Hamiltonian and momentum constraints, respectively. Below they will be called the lapse function and shift vector constraints in the Lagrangian method. 
As will be detailed shortly, in a flat background the following gauge is suitable for quantization 
\bea
n=1 \quad,\quad N_m=0  \la{gf}
\eea 
With $n=1$, the shift vector constraint is automatically satisfied and the Hamiltonian density takes
\bea
{\mathscr H}_{gauge\; fixed} &=& -\sqrt{-h}\: {\cal C}=-\sqrt{-h}\;(\cR-K^2+K_{mn}K^{mn})
\eea
where ${\mathscr H}_{gauge\;fixed}$ denotes the gauge-fixed Hamiltonian density.
The lapse function constraint (to be discussed shortly in \rf{NnconH} below) with the gauge-fixing \rf{gf} implies
\bea
H_{gauge\;fixed}|phys>=0  \la{Hconstr}
\eea 
where $H_{gauge\;fixed}$ denotes the gauge-fixed Hamiltonian.
It is in the full nonlinear sense. 
Note the dual roles of the gauge-fixed Hamiltonian: it is the operator that governs the $x^3$-evolution and at the same time acts as the lapse function constraint. 
In \cite{Park:2015ybl} (see also \cite{Higuchi:1991tk} for an earlier discussion), the condition above has been interpreted as to imply that the physical states must be contained in a hypersurface in the asymptotic region.\footnote{A complementary mathematical discussion based on jet bundle theory \cite{Saunders} can be found in \cite{Park:2014qoa,Park:2015qxa}.} 
It is this reduction that allows a description of the 4D physics through the 3D window. See \cite{York:1972sj,Moncrief:1989dx,Fischer:1996qg,Marsden,Gay-Balmaz:2014ena,deBoer:2003vf,Gomes:2010fh} for various reduction-related works.

The details of the gauge-fixing procedure are as follows. For a Minkowski background, ${x^3}$ can then be taken as one of the spatial coordinates. 
Although above, we have not distinguished the fluctuations from the backgrounds, we will do so from now on:
\bea
\mbox{field}=\mbox{background}+\mbox{fluctuation}
\eea 
and the background will be indicated by a subscript zero.
The flat background is such that $n_0=1$ and $N_0^m=0$ where the subscript zero indicates that the field is the background - as opposed to the fluctuation - quantity.  
By using the gauge symmetry, the fluctuation part of the lapse function can be gauged away leaving it fixed to its background value:
\bea
n=n_0
\eea
We also adopt the synchronous-type gauge-fixing:
\bea
 N_m=0
\eea
Since the flat background has a vanishing shift vector, this fixing corresponds to gauging away the fluctuation part. 
For a flat background this gauge can definitely be chosen. For a Schwarzschild background for example, the steps so far are quite similar other than that $x_3$ is taken as the radial direction and $n_0=\Big(1-\fr{2GM}{r}\Big)^{-1}$. One of the potential obstacles for applying the present quantization method to a more general, say, time- and position- dependent background is the issue of whether or not this shift vector-fixing is available; we will come back to this issue toward the end.

The induced shift vector constraint, which is nothing but the shift vector field equation,
\bea
\N^n(-K_{mn}+K h_{mn})=0,
\eea
which translates into
\bea
\nabla_a n=0,  \la{sc2}
\eea
is automatically satisfied with the gauge-fixing above, $n_0=1$. Whether this would remain valid for a more general background is another potential obstacle for quantization in that background to be discussed below. 
We impose the lapse function constraint (i.e., the field equation of the lapse field) in its weaker form as the physical state condition
\bea
\Big[\cR-K^2+K_{mn}K^{mn}\Big]|phys>=0 
\la{NnconH}
\eea
This is a part of the quantization proposal; an analogous step, i.e., the gauge-fixing-induced constraint is well-established in string theory and leads to the Virasoro constraint.

In the next section, we will extend the reduction scheme to an arbitrary asymptotically flat background. Again, the gauge-fixing is a critical initial step for the reduction. Let us note that $n_0=1$ for a flat background, so the condition \rf{sc2} is trivially satisfied. 
Similarly, the Schwarzschild background in the Schwarzschild coordinates satisfies the condition. As a matter of fact, the stipulation \rf{sc2} can be made satisfied by choosing the following gauge \cite{Landau,Shapiro,Gourgoulhon}
\bea
n=1\quad ,\quad N_m=0 \la{gs}
\eea 
The real question is whether or not this gauge-fixing can always be chosen and things would remain fine. We believe this is a matter of subtleties not entirely sorted out in the literature. In general, a gauge must be chosen in a manner not interfering with the physics that one intends to study. Also (perhaps not unrelated), there is no such gauge that is valid for all physical configurations, the so-called Gribov-Singer problem \cite{Isham}. (See, e.g., \cite{Eichhorn:2013ug} for a recent discussion.) What must still be true is that there should exist an (at least locally) valid choice of a gauge for a given class of backgrounds. Practically and also in the present context, this means that only the backgrounds, including time- and position- dependent ones, that are compatible with the gauge \rf{gs} will be considered, and this should be sufficient for our purpose. 
With the condition for the reduction settled, we now turn to the renormalizability.

\subsection{time- and position- dependent backgrounds}

For renormalizability the reduction is not the full story: the behavior of $K_{ab}$ at the asymptotic region should be constrained as well. The constraint is mild enough to be guaranteed by the asymptotic flatness. The bulk and hypersurface quantities are related by the following relations \cite{Aliev:2004ds}:
\bea
R_{mrpq}&=&\cR_{mrpq}+K_{mq}K_{rp}-K_{mp}K_{rq} \\
R_{3mrp}&=&N^l(\cR_{lmrp}+K_{lp}K_{mr}-K_{lr}K_{mp})
           -n(\nabla_r K_{mp}-\nabla_p K_{mr})\nn\\
R_{m3p3}&=&N^l\Big[N^r(\cR_{rmlp}+K_{rp}K_{ml}-K_{rl}K_{mp})
           -n(\nabla_l K_{mp}-\nabla_p K_{ml})\Big]  \nn\\
        &&  -n\Big(\mathscr{L}_{(\pa_{x^3}-N^q\pa_q)}K_{mp}+\nabla_m \nabla_p n\Big)            
            +n\Big(N^l(\nabla_m K_{lp}-\nabla_l K_{mp})+n K_{mr}K^r_{p}\Big);\nn
\eea
\bea
R_{mr}&=&\cR_{mr}-\fr1{n}\Big(\mathscr{L}_{(\pa_{x^3}-N^q\pa_q)}K_{mr}+\nabla_m \nabla_r n\Big)
         -KK_{mr}+2K_{ml}K^l_r  \nn\\
R_{m3}&=& N^l\Big[\cR_{ml}-\fr1{n}\Big(\mathscr{L}_{(\pa_{x^3}-N^q\pa_q)}K_{ml}+\nabla_m         
                 \nabla_l n\Big)-KK_{ml}+2K_{mr}K^r_l\Big]-n(\nabla_m K-\nabla_lK^l_m)\nn\\                
R_{33}&=&N^mN^r\Big[\cR_{mr}-\fr1{n}\Big(\mathscr{L}_{(\pa_{x^3}-N^q\pa_q)}K_{mr}
           +\nabla_m \nabla_r n\Big)-KK_{mr}+2K_{ml}K^l_r \Big]\nn\\
       &&  -n(\mathscr{L}_{\pa_{x^3}}K+\nabla_l\nabla^l n)  
       -n^2 K_{lr}K^{lr}+2n N^r \nabla^l\Big(K_{lr}-\fr12 h_{lr} K\Big)
\eea
and
\bea
R=\cR-K_{mr}K^{mr}-K^2-\fr2{n}\Big(\mathscr{L}_{(\pa_{x^3}-N^q\pa_q)}K+\nabla_m \nabla^m n\Big)
\eea
With the gauge-fixing \rf{gs}, one gets, for the Riemann tensor, 
\bea
R_{mnpq}&=&{\cal R}_{mnpq}+K_{mq}K_{np}-K_{mp}K_{nq} \nn\\
R_{3mnp}&=&-n_0(\nabla_n K_{mp}-\nabla_p K_{mn})\nn\\
R_{m3p3}&=&
          -n_0\mathscr{L}_{\pa_{r}}K_{mp}            
            +n_0^2 K_{mr}K^r_{p}   \la{rrt}
\eea
As we will discuss in detail in section 3, we impose the Neumann boundary condition. The Neumann boundary condition with the reduction should imply that the second fundamental form at the boundary vanishes:  
\bea
K_{mn} \ra  0
\eea
which is assured by the asymptotically flatness. With this, the renormalization procedure parallels with the flat case.

After all, the analysis in this section shows that the present findings are consistent with the common lore regarding renormalizability as stated in the beginning of section 2. This seems natural: the renormalization procedure in an asymptotically flat background parallels that of a flat spacetime since the physical degrees of freedom are associated with the hypersurface at the asymptotic region where the background spacetime becomes flat.

\subsection{ramifications}

With the tasks in the previous subsections completed, we muse over various implications of the reduction. This will allow us to set the stage for building a global picture of the scattering and BHI, the main topic for the next section. 

As proposed in \cite{Park:2015ybl} we take the Wheeler-DeWitt equation \rf{Hconstr} as to imply the reduction of the physical states to the holographic screen. This identification raises the issue of how to handle the boundary conditions and its dynamics.  
There are several indications that the Neumann boundary condition is the right choice. Since the physical states are reduced to the boundary surface, one needs a boundary condition that gives dynamics to the surface; otherwise the system will have only the non-perturbative configurations. The fact that the radial coordinate has played the role of time seems to suggest that one may impose the `$r$-Dirichlet' boundary condition, namely the Neumann boundary condition. (A recent analysis on the Neumann boundary condition can be found in \cite{Krishnan:2016mcj}.) In fact, the Neumann boundary condition makes $K_{mn}$ non-dynamic, a necessary condition for the renormalizability. Then the asymptotic flatness makes it vanish at the asymptotic region.

One of the main objectives of the (3+1)-splitting is to separate the physical degrees of freedom. With the Neumann boundary condition the hypersurface is dynamical; the next step is to analyze the genuine-time dynamics of the 3D hypersurface.  
 Although the formalism is not fully covariant, it has certain advantages (such as less formalism and mathematical machinery) over the covariant approach yet to be developed just as the lightcone quantization of string has certain advantages over the more sophisticated BRST quantization.

All of these ideas will be illustrated in the next section where we construct the setup for BHI with the Schwarzschild case. One of the central features of the setup is the enlarged Hilbert space of \cite{Freidel:2016bxd,Donnelly:2016auv}, perhaps a little more systematized and generalized below. At a supposedly elementary level, it offers a hint at how to tackle an aspect of BHI that we find puzzling: how come a Schwarzschild observer observes the black hole evaporation since the Schwarzschild coordinates are well-behaved outside the horizon where the mass must thus be a conserved quantity? The evaporation seems to suggest that the mass is not a conserved quantity. The Neumann boundary condition provides a direct answer to the question; the rationale for its relevance can be found in the enlarged Hilbert space. At a more sophisticated level, the enlarged Hilbert space provides a stage to realize the generalized spontaneous symmetry breaking and Goldstone degrees of freedom.

\section{Scattering, vacuum transition and BHI}

With the quantization scheme extended to include a larger class of backgrounds, let us take up the black hole information problem.  
One of the things that make the BHI analysis much subtler than otherwise is that in order to properly describe the states, one must carefully consider, among other things, the boundary dynamics. 
The fact that the physical degrees of freedom are associated to the hypersurface at the asymptotic region calls for a refined formulation of the information and information paradox.

According to the AdS/CFT-type dualities, the bulk physics can be described by the degrees of freedom at the boundary. Although the correspondence has been well established by now, it has a peculiar aspect regarding the boundary conditions. Namely, if one imposes a Dirichlet boundary condition - which one normally does - one must worry about the possibility of contradiction: the dual boundary theory is dynamical whereas the Dirichlet boundary condition dictates non-dynamism of the boundary degrees of freedom.      
As we will see, the problem does not arise once one chooses a Neumann boundary condition.

Below in section 3.1, we start by presenting the ``big" picture. We first give a detailed construction of the setup, which is the system consisting of the bulk states with the boundary degrees of freedom embedded. We often implicitly turn to the SQSD.
After generalities, we take in section 3.2 the example of a Schwarzschild black hole to illustrate how to implement the general ideas. Two main tasks are: firstly, we carry out the gauge-fixing-induced reduction scheme by employing the tool of dimensional reduction to the hypersurface of foliation \cite{Park:2013iqa,Park:2013vpa,Park:2013bma}, a variant of the Kaluza-Klein procedure (see, e.g.,\cite{VanNieuwenhuizen:1981ae,Duff:1986hr} for reviews of Kaluza-Klein compactification). Secondly, we present the global formulation of the BHI with the bulk and embedded boundary degrees of freedom.
In section 3.3, we elaborate on and quantify the view put forth in \cite{Park:2013rm} that there should exist two different kinds of  `radiations,' one information-carrying and the other information-neutral. This view is now founded on the fact that there exists a hierarchy in the excitations and thus in their transitions. The scattering of perturbative boundary states must be associated with the information-carrying radiation. For the information-neutral `radiation,' we discuss the process for a Schwarzschild black hole to decay into a Minkowski vacuum through the vacuum transition. The bubble diagrams will be responsible for the disintegration; we propose that they should be responsible for the information-neutral `radiation.' The picture also suggests that it is this information-neutral `radiation' that should be identified with the Hawking radiation.

\subsection{scattering around  black hole: generalities}

For the global perspective of the BHI problem, it is important to note that the system has the `embedded' boundary degrees of freedom as well as the offshell bulk degrees of freedom in the sense detailed in section 2. 
For an illustration of the idea, let us consider the 5D gravity theory considered in \cite{Sato:2002kv}\cite{Hatefi:2012bp} (see, e.g., \cite{Niarchos:2015moa,Grignani:2016bpq,Maxfield:2016vpw} for related works) that was obtained by compactifying type IIB supergravity on $S^5$. The following was shown. One sets out to find a solution to the system by employing the Hamilton-Jacobi procedure. Instead of finding a single solution, one obtains a class of solutions with the ``moduli fields." A close inspection of the constraints of the mother 5D theory reveals that the moduli fields are nothing but the worldvolume gauge field \cite{Hatefi:2012bp}. 
The worldvolume gauge theory will have two types of excitations: the solutions to its field equations and Fock space excitations around that solution. In other words, the gauge theory has perturbative and non-perturbative excitations. 
Now consider the original 5D gravity theory and its path integral. The fact that it is non-renormalizable is irrelevant for our purposes. The analysis along the line of  \cite{Park:2014tia} will reveal that the physical degrees of freedom should be those of the gauge theory (although this identification will be less direct due to the involvement of the Hamilton-Jacobi equation) and thus they must be the main focus for the scattering process taking place in the bulk. The analyses of \cite{Sato:2002kv} and \cite{Hatefi:2012bp} explicitly show how the dual boundary degrees of freedom arise from the bulk theory through the Hamilton-Jacobi equation.\footnote{At a deeper level, it should be a ``spontaneous symmetry breaking" that is at work behind \cite{Park:2008sg,Park:2013bma,Hatefi:2012bp} and the YM field can be viewed as to represent the `Goldstone' particles. \la{ssb}} Also, the framework naturally suggests that there will be different kinds of the excitations and transitions for the 5D theory.

Since worldvolume gauge degrees of freedom appear from the original gravity theory through the intricate Hamilton-Jacobi method, making a tangible connection between the boundary theory and the bulk behavior cannot but be complicated and less direct. Unlike the 5D example above, the foliation-based reduction mechanism of \cite{Park:2013iqa,Park:2014tia,Park:2015ybl} makes it possible to make an intuitive and direct connection between the boundary degrees of freedom and the boundary-dynamics-induced bulk response. 
There are two different kinds of physical excitations of the bulk-hypersurface combined system that we mainly focus on in this work. The first kind is the classical solutions of the bulk field equations. For example, the Schwarzschild solution is a higher energy solution of the vacuum Einstein equation than a Minkowski spacetime. The second kind is the ``fine" excitations specified by hypersurface theory creations operators around a given classical solution.\footnote{In general, there will be a third kind: the non-perturbative solutions of the boundary theory, as we have seen in the example of 5D gravity above. For simplicity of the discussion, they will be set aside.} 

The bulk dynamics is responsible for the vacuum transition through non-perturbative dynamics which cannot be captured by the boundary theory.  
The bulk transition under consideration is the decay, for instance, from a Schwarzschild black hole configuration to a Minkowski configuration. Although the evaporation of a black hole is a widely accepted picture, one may wonder whether such a transition could occur given that the Schwarzschild black hole should be an energy eigenstate (which we will shortly argue should not be true generically) as well as what the mechanism behind such a transition would be. We present the answers to these questions as advocated by the quantization scheme.



\subsection{example}

To prevent things from getting too abstract, let us illustrate the ideas by constructing the scattering setup in the case of a Schwarzschild black hole,  
\bea
ds^2=-\Big(1-\fr{2GM}{r}\Big)dt^2+\Big(1-\fr{2GM}{r}\Big)^{-1}dr^2+r^2(d\th^2+\sin^2\th d\varphi^2)
\eea
Setting apart the irrelevant boundary terms, the ADM form of the action is 
\bea
S = \int d^4 x\;n\sqrt{-h}\; \Big[\;\cR+K^2-K_{mn}K^{mn}\;\Big]
\eea
The first step of constructing the setup of the entire bulk and boundary system is to reduce this action to a hypersurface at a fixed $r$ that will be taken at $r=\infty $ at the end. The detailed steps of the reduction are more complicated than the Minkowski case and can be found in \cite{Park:2015xoa}. Let us review the salient features of the procedure. 
In addition to the gauge-fixings of the lapse function and shift vector, which leads to
\bea
{\mathscr H}=-\sqrt{-h}\;(\cR-K^2+K_{mn}K^{mn})
\eea
the trace piece of the 4D metric must be gauged away, and this renders $K$ non-dynamical. In other words, the field $K$ gets to be set to its background value:
\bea
K=K_0;
\eea  
this term thus becomes a cosmological constant (or ``function," more precisely) term. 
The lapse function and gauge-fixed Hamiltonian have the same form. As discussed in \cite{Park:2015ybl} (and in an earlier linear-level analysis in \cite{Higuchi:1991tk}) this has been interpreted as to imply that the physical states of the theory are contained in the hypersurface at the asymptotic region. More explicitly, we take the following form of the metric reduction: 
\bea
h_{mn} = h_{0mn}(r)+\tilde{h}_{mn}(t,\th,\f) \equiv \g_{mn}(t,r,\th,\f) \la{hred}
\eea
where $h_{0mn}(r)\equiv diag\Big(-1+\fr{2GM}{r},r^2,r^2\sin^2\th\Big)$, and $\tilde{h}_{mn}(t,\th,\f)$ denotes the fluctuation metric around the fixed-$r$ hypersurface. We have introduced another notation, $\g_{mn}$, to stress the split form of the 3D metric.

Several remarks are in order. The asymptotic flatness concerns the background metric but not the fluctuation. The fluctuation $\tilde{h}_{mn}(t,\th,\f)$, which would get removed by the Dirichlet boundary condition in the conventional treatment, now survives and governs the hypersurface dynamics. It can be viewed as the `Goldstone' degrees of freedom (see footnote \ref{ssb}).  The presence of such fluctuating fields means that the bulk system is not isolated and the mass of the black hole is not a conserved quantity with the Neumann boundary condition, which is consistent with the black hole's decay. 

The circumstance seems to suggest that the Hilbert space needs to be enlarged by including all of the possible boundary conditions, Dirichlet and Neumann (and more general boundary conditions as well in general).\footnote{To a certain extent, this is already being practiced in the literature. 
This is basically the situation associated, e.g., with the Hartle-Hawking and Boulware vacua: those vacua are associated with the two different slicings, i.e., two different foliations of the same bulk geometry. In both coordinates Dirichlet boundary conditions are imposed but the boundary surfaces are different.} 
The Dirichlet boundary condition will represent a measure zero state, $\tilde{h}_{mn}(t,\th,\f)=0$, in the multitude of the possible boundary conditions. The disintegrating black hole should feed the hypersurface dynamics.


The `largely' 4D covariant path integral approach of \cite{Park:2016zgt,Park:2015ota} can now be followed.\footnote{One may instead adopt the entirely-3D onshell operator approach of \cite{Park:2014noa}.
Collecting all of the above, the reduced action can be written
\bea
S_{HS} = \int dt d\th d\f\;n_0\sqrt{-\g}\; \Big[\;\cR(\g_{ab})+K_0^2-K_{0mn}K_{0pq}\;\g^{mp}\g^{nq}\;\Big] \la{3Dact}
\eea 
where the subscript `HS' denotes `hypersurface,' and
\bea
n_0^2= \Big(1-\fr{2GM}{r}\Big)^{-1},\quad
K_{0mn}&=&\fr1{2n_0}\pa_{{r}} h_{0mn} \;,\quad
K_0= h_0^{mn}K_{0mn}
\eea
The coordinate $r$ has been `demoted' to a parameter.
With this action of the dynamical field $\g_{ab}$ and the onshell operator quantization, one can study the dynamics on the hypersurface. This method is analogous to the lightcone quantization of string theory whereas the largely covariant approach is analogous, to a certain extent, to the old covariant quantization.
} 
The approach is suitable for capturing the nonperturbative bulk vacuum transition as well. In this approach, one would obtain the 4D covariant effective action first and then reduce it by taking the external states of the Feynman diagrams to be the physical states. This could be implemented by substituting \rf{hred} with the gauge condition \rf{gs} into the 4D covariant offshell effective action.

\begin{figure}[tbp]
\centering 
\includegraphics[width=1.0\textwidth,trim=0 480 0 40,clip]{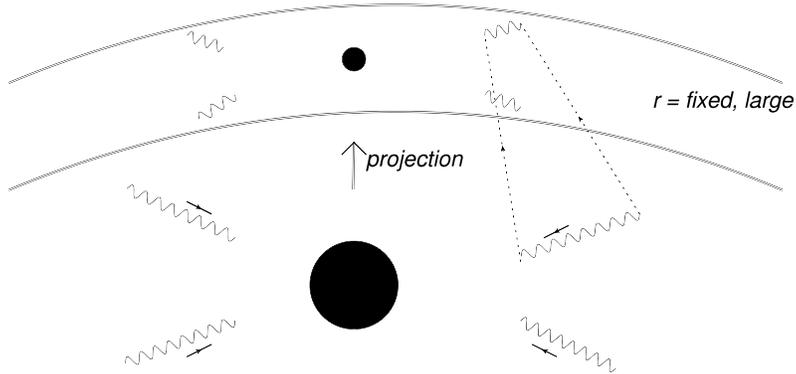}
\caption{\label{fig:i}  4D scattering around BH projected onto 3D hypersurface}
\end{figure}

\subsection{vacuum decay via Hawking radiation}

Let us take a pause and have an overview of what's been discussed so far. We have started with the 4D Einstein-Hilbert action. Through the ADM Hamiltonian and Lagrangian analysis, the reduction of the physical states onto the hypersurface in the asymptotic region has been delivered.   
In the entirely 3D onshell operator approach \cite{Park:2014noa}, one would use the action given in \rf{3Dact} to proceed with the quantization and dynamics on the hypersurface. The disadvantage of this approach is that it cannot capture the bulk effects such as the vacuum transition. 
In the alternate largely covariant approach \cite{Park:2016zgt}, one may path-integrate over the bulk modes and obtain the 4D offshell effective action.  Restricting the external states to the physical states, the 4D offshell action can be reduced to the 3D effective action. With this approach, the non-perturbative vacuum transition can be dealt with (more below), which will be an important part of the BHI analysis.
A cartoon image of the largely covariant method is depicted in Fig. 1.



At this point let us take a detour to the recent series of works that hinge on the asymptotic symmetry group, the BMS group \cite{Geroch:1972up,Ashtekar:1978zz,Arcioni:2003td,Carney:2017jut,Strominger:2017aeh,Bousso:2017dny}. We will shortly resume the discussion above and compare these works with our view. In \cite{Strominger:2017aeh}, a certain subgroup of the BMS group has been proposed to be the hair of the black hole. The pivot of the argument is as follows. The spontaneous symmetry breaking of the subgroup has been found responsible for generation of the soft particles such as soft gravitons. Since the soft particles are infinite in number, they may carry most of the information that is not apparent on the black hole or the outgoing hard radiation.


We take the presence of such hair as an indication of the existence of the boundary's own dynamics \cite{Park:2016fxc}.
To make a more detailed comparison of our approach with these works, let us give some further thought to the two examples in the previous subsections, the 5D gravity theory and pure Einstein gravity. As a matter of fact, the two examples can be viewed as a procedure of solution-finding and generalized spontaneous symmetry breaking. It is easier to see this with the 5D example. A solution of worldvolume gauge field equation, once embedded into the Kaluza-Klein ansatz, in turn yields a solution of the original 5D gravity. The gauge field can be viewed as the Goldstone degrees of freedom.
Similarly, gauge-fixing-induced reduction procedure of the 4D Einstein gravity example can be viewed as a solution finding procedure. After reduction to 3D, the system admits the 3D ``Schwarzschild black hole" \cite{Park:2013iqa} as a solution. (In other words, \rf{3Dact} admits, by design, $\g_{mn}=0$ as a solution.) The dynamical field $\tilde{h}_{mn}(t,\th,\f)$ in \rf{hred} is the fluctuation around that 3D solution and plays the role of the Goldstone field; the symmetry spontaneously broken is the 3D diffeomorphism of the hypersurface.


Since the vacuum transition is involved in the whole-nine-yards setup of BHI, it is necessary to employ a generalized version of the techniques widely used in the physics of the vacuum bubble creation in cosmology \cite{Callan:1977pt,EWeinberg}. To be able to apply those techniques, one of the key ingredients is a solution that interpolates the Minkowski vacuum and Schwarzschild black hole (say, assuming the presence of matter). The path integral can be evaluated perturbatively: it will be the sum of the vacuum bubble diagrams. 
 The explicit quantitative analysis is expected to be exceedingly involved. For one thing, (existence of) the interpolating solution is not known. Before we outline the necessary steps - which would be useful in the actual analysis, given its complexity - let us state the anticipated outcomes first since there are clearly anticipated outcomes in spite of this complexity. The path integral calculation for the transition amplitude is essentially the effective action computation, and the resulting quantum action will contain the vacuum decay effect in it. The action can then be further analyzed to study the hypersurface dynamics at the 3D quantum level once the 3D modes are path-integrated. After the 3D path integral, the resulting effective action can be used to study the perturbative scattering process in the background of the information-neutral `radiation' associated with the bubble diagrams. 
To our view it should be the information-neutral radiation emerging from the vacuum transition that should be taken as the Hawking radiation.\footnote{In \cite{Park:2013rm}, the process of flux loss and black hole mass increase was discussed. The mass-gaining process must be in reverse to the vacuum transition.}

Let us now get to the more technical side of the story, the actual setup of the bulk and boundary path integral. Although matter fields must be present to be realistic, we focus on the gravity sector for simplicity. At the superficial level, the path integral to be evaluated is the usual one:
\bea
\int Dg_{\m\n}\;e^{iS}
\eea 
where the action $S$ has the Einstein-Hilbert part \rf{eha} for the gravity sector. Here come the complications: split the metric into the interpolating solution and fluctuation, and integrate out the bulk modes to obtain the offshell 4D effective action (see \cite{Park:2016zgt,Park:2017dib} for an illustration). This step should then accommodate the vacuum transition effect between the two vacua. The 4D effective action can be reduced to 3D once the metric is reduced by following the steps analogous to those presented in section 3.2. 
Once the 3D action is obtained, one may perform the 3D hypersurface path integral to account for the 3D dynamics. Afterwards one will obtain the complete quantum-corrected action. 
The physical picture is that the whole process will be superposition of the perturbative scattering and non-perturbative vacuum transition: the perturbative scattering will take place on the background of vacuum transition that turns up through the creation of bubbles and information-neutral (or information-minimal) radiation.
Not only the boundary dynamics but also the bulk physics should contain the information: the solution of the fully quantum effective action will give the quantum deformed geometry that reflects part of the information \cite{Park:2017dib}.
The information-neutral radiation will interact, on its way out, with the horizon vicinity and while doing so the characteristics of the vicinity, i.e., the information coded in the vicinity,  would be revealed to a distant observer.


\section{Conclusion}

One of the lessons emerging from the various recent studies is that the boundary conditions and boundary dynamics 
are of central importance in quantizing gravity as well as in black hole information. In this work, they have played a critical role in quantization of gravity in an arbitrary asymptotic background. We have pointed out that the enlarged Hilbert space proposed in \cite{Freidel:2016bxd,Donnelly:2016auv} or the generalization thereof is tied with the hypersurface dynamics. 
After extending the quantization scheme of \cite{Park:2014tia} to a much larger class of backgrounds, i.e., arbitrary asymptotic flat backgrounds, we have put forth a coherent and global framework suitable for studying the black hole information and related problems. 
We have applied the scheme to construct the setup for scattering of the physical states that have support at the boundary hypersurface. The quantization scheme suggests a solution of the black hole information paradox. To be specific, we have considered a Schwarzschild background.
We have noted the hierarchy in the transitions: the transitions among different Fock states and vacuum transition.  
The `hair' in part is associated with the boundary's own dynamics. We have proposed that the information-neutral radiation from the vacuum transition be associated with the Hawking radiation. 

\vspace{.2in}
There are several interesting future directions:
\vspace{.1in}

In section 3, the basic conceptual framework for the BHI and paradox has been laid out. The outlined steps should be explicitly carried out at some point. The highest hurdle will be to obtain the interpolating solution. Constructing the Green's function will be technically involved too but it should be possible to employ a certain approximation method. If one's focus is on the vacuum transition, the AdS case analyzed in \cite{Park:2016fxc,Park:2016vam} should presumably serve the purpose better.   

There may be some relation between the BMS group and 3D diffeomorphism of the hypersurface, although the fact that in our case, the split-out direction is one of the spatial directions would make the comparison less simple. 
The 3D diffeomorphism of the action \rf{3Dact} - which is implicitly expanded around the `3D Schwarzschild,' which is $(R\times S^2)$ topologically - will be broken to the symmetry group of this background. It will be of some interest to explore whether one can come up with a more sophisticated 4D covariant approach along the lines of the asymptotic symmetries. 

Progress on some of these issues will be reported in the future.




\newpage
\appendix

\renewcommand{\theequation}{A.\arabic{equation}}
\setcounter{equation}{0}

%
%
%
%

\newpage


\begin{thebibliography}{99}




\bibitem{DeWitt:1975ys} 
B.~S.~DeWitt,
``Quantum Field Theory in Curved Space-Time,''
Phys.\ Rept.\  {\bf 19}, 295 (1975).
doi:10.1016/0370-1573(75)90051-4



\bibitem{'tHooft:1974bx}
G.~'t Hooft and M.~J.~G.~Veltman,
``One loop divergencies in the theory of gravitation,''
Annales Poincare Phys.\ Theor.\ A {\bf 20}, 69 (1974).


\bibitem{Brill:1970df} 
  D.~R.~Brill and R.~H.~Gowdy,
  ``Quantization of general relativity,''
  Rept.\ Prog.\ Phys.\  {\bf 33}, 413 (1970).
  doi:10.1088/0034-4885/33/2/301



\bibitem{Stelle:1976gc}
K.~S.~Stelle,
``Renormalization of Higher Derivative Quantum Gravity,''
Phys.\ Rev.\ D {\bf 16} (1977) 953.



\bibitem{Antoniadis:1986tu} 
I.~Antoniadis and E.~T.~Tomboulis,
``Gauge Invariance and Unitarity in Higher Derivative Quantum Gravity,''
Phys.\ Rev.\ D {\bf 33}, 2756 (1986).





\bibitem{Weinberg3} 
S. Weinberg, in ``General Relativity, an Einstein Centenary Survey,"
edited by S. Hawking and W. Israel, Cambridge (1979)



\bibitem{Reuter:1996cp} 
M.~Reuter,
``Nonperturbative evolution equation for quantum gravity,''
Phys.\ Rev.\ D {\bf 57}, 971 (1998)
[hep-th/9605030].

\bibitem{Odintsov:1991yx} 
  S.~D.~Odintsov,
  ``Does the Vilkovisky-De Witt effective action in quantum gravity depend on the configuration space metric?,''
  Phys.\ Lett.\ B {\bf 262}, 394 (1991).
  doi:10.1016/0370-2693(91)90611-S


\bibitem{Barvinsky:1993zg}
A.~O.~Barvinsky, A.~Y.~Kamenshchik and I.~P.~Karmazin,
``The Renormalization group for nonrenormalizable theories: Einstein gravity with a scalar field,''
Phys.\ Rev.\ D {\bf 48} (1993) 3677
doi:10.1103/PhysRevD.48.3677
[gr-qc/9302007].


\bibitem{Carlip:2001wq} 
S.~Carlip,
``Quantum gravity: A Progress report,''
Rept.\ Prog.\ Phys.\  {\bf 64}, 885 (2001)
[gr-qc/0108040].





\bibitem{Ambjorn:2012jv} 
J.~Ambjorn, A.~Goerlich, J.~Jurkiewicz and R.~Loll,
``Nonperturbative Quantum Gravity,''
Phys.\ Rept.\  {\bf 519}, 127 (2012)
doi:10.1016/j.physrep.2012.03.007
[arXiv:1203.3591 [hep-th]].




\bibitem{Woodard:2014jba} 
R.~P.~Woodard,
``Perturbative Quantum Gravity Comes of Age,''
arXiv:1407.4748 [gr-qc].


\bibitem{Donoghue:2015hwa} 
J.~F.~Donoghue and B.~R.~Holstein,
``Low Energy Theorems of Quantum Gravity from Effective Field Theory,''
J.\ Phys.\ G {\bf 42}, no. 10, 103102 (2015)
doi:10.1088/0954-3899/42/10/103102
[arXiv:1506.00946 [gr-qc]].






\bibitem{Hawking:1976ra} 
S.~W.~Hawking,
``Breakdown of Predictability in Gravitational Collapse,''
Phys.\ Rev.\ D {\bf 14}, 2460 (1976).
doi:10.1103/PhysRevD.14.2460


\bibitem{Page:1993up} 
D.~N.~Page,
``Black hole information,''
hep-th/9305040.


\bibitem{Mathur:2009hf} 
  S.~D.~Mathur,
  ``The Information paradox: A Pedagogical introduction,''
  Class.\ Quant.\ Grav.\  {\bf 26}, 224001 (2009)
  doi:10.1088/0264-9381/26/22/224001
  [arXiv:0909.1038 [hep-th]].

\bibitem{Hooft:2016vug} 
  G.~'t Hooft,
  ``The firewall transformation for black holes and some of its implications,''
  arXiv:1612.08640 [gr-qc].

\bibitem{Polchinski:2016hrw} 
  J.~Polchinski,
  ``The Black Hole Information Problem,''
  arXiv:1609.04036 [hep-th].





\bibitem{Park:2014tia} 
I.~Y.~Park,
``Hypersurface foliation approach to renormalization of ADM formulation of gravity,''
Eur.\ Phys.\ J.\ C {\bf 75}, no. 9, 459 (2015)
doi:10.1140/epjc/s10052-015-3660-x
[arXiv:1404.5066 [hep-th]].





\bibitem{Park:2015ybl} 
  I.~Y.~Park,
  ``Reduction of gravity-matter and dS gravity to hypersurface,''
  Int.\ J.\ Geom.\ Meth.\ Mod.\ Phys.\  {\bf 14}, no. 06, 1750092 (2017)
  doi:10.1142/S021988781750092X
  [arXiv:1512.08060 [hep-th]].



\bibitem{Park:2014noa} 
  I.~Y.~Park,
  ``Lagrangian constraints and renormalization of 4D gravity,''
  JHEP {\bf 1504}, 053 (2015)
  doi:10.1007/JHEP04(2015)053
  [arXiv:1412.1528 [hep-th]].



\bibitem{Park:2015xoa} 
I.~Y.~Park,
``Holographic quantization of gravity in a black hole background,''
J.\ Math.\ Phys.\  {\bf 57}, no. 2, 022305 (2016)
doi:10.1063/1.4942101
[arXiv:1508.03874 [hep-th]].



\bibitem{Park:2016zgt} 
  I.~Y.~Park,
  ``One-loop renormalization of a gravity-scalar system,''
  Eur.\ Phys.\ J.\ C {\bf 77}, no. 5, 337 (2017)
  doi:10.1140/epjc/s10052-017-4896-4
  [arXiv:1606.08384 [hep-th]].







\bibitem{Park:2013rm} 
I.~Y.~Park,
``On the pattern of black hole information release,''
Int.\ J.\ Mod.\ Phys.\ A {\bf 29}, 1450047 (2014)
doi:10.1142/S0217751X1450047X
[arXiv:1301.6320 [hep-th]].









\bibitem{Benguria:1976in} 
  R.~Benguria, P.~Cordero and C.~Teitelboim,
  ``Aspects of the Hamiltonian Dynamics of Interacting Gravitational Gauge and Higgs Fields with Applications to Spherical Symmetry,''
  Nucl.\ Phys.\ B {\bf 122}, 61 (1977).
  doi:10.1016/0550-3213(77)90426-6

\bibitem{Witten:1988hf} 
  E.~Witten,
  ``Quantum Field Theory and the Jones Polynomial,''
  Commun.\ Math.\ Phys.\  {\bf 121}, 351 (1989).
  doi:10.1007/BF01217730

\bibitem{Balachandran:1991dw} 
  A.~P.~Balachandran, G.~Bimonte, K.~S.~Gupta and A.~Stern,
  Int.\ J.\ Mod.\ Phys.\ A {\bf 7}, 4655 (1992)
  doi:10.1142/S0217751X92002106
  [hep-th/9110072].

\bibitem{Smolin:1995vq} 
  L.~Smolin,
  ``Linking topological quantum field theory and nonperturbative quantum gravity,''
  J.\ Math.\ Phys.\  {\bf 36}, 6417 (1995)
  doi:10.1063/1.531251
  [gr-qc/9505028].
 
  

\bibitem{Parattu:2016trq} 
  K.~Parattu, S.~Chakraborty and T.~Padmanabhan,
  ``Variational Principle for Gravity with Null and Non-null boundaries: A Unified Boundary Counter-term,''
  Eur.\ Phys.\ J.\ C {\bf 76}, no. 3, 129 (2016)
  doi:10.1140/epjc/s10052-016-3979-y
  [arXiv:1602.07546 [gr-qc]].



\bibitem{Krishnan:2016mcj} 
C.~Krishnan and A.~Raju,
``A Neumann Boundary Term for Gravity,''
arXiv:1605.01603 [hep-th];
 C.~Krishnan, K.~V.~P.~Kumar and A.~Raju,
  ``An alternative path integral for quantum gravity,''
  JHEP {\bf 1610}, 043 (2016)
  doi:10.1007/JHEP10(2016)043
  [arXiv:1609.04719 [hep-th]].




\bibitem{Lehner:2016vdi} 
L.~Lehner, R.~C.~Myers, E.~Poisson and R.~D.~Sorkin,
``Gravitational action with null boundaries,''
arXiv:1609.00207 [hep-th].


\bibitem{Baierlein:1962zz} 
  R.~F.~Baierlein, D.~H.~Sharp and J.~A.~Wheeler,
  ``Three-Dimensional Geometry as Carrier of Information about Time,''
  Phys.\ Rev.\  {\bf 126}, 1864 (1962).
  doi:10.1103/PhysRev.126.1864





\bibitem{Park:2016fxc} 
  I.~Y.~Park,
  ``Quantum "violation" of Dirichlet boundary condition,''
  Phys.\ Lett.\ B {\bf 765}, 260 (2017)
  doi:10.1016/j.physletb.2016.12.026
  [arXiv:1609.06251 [hep-th]].













\bibitem{Freidel:2016bxd} 
  L.~Freidel, A.~Perez and D.~Pranzetti,
  ``Loop gravity string,''
  Phys.\ Rev.\ D {\bf 95}, no. 10, 106002 (2017)
  doi:10.1103/PhysRevD.95.106002
  [arXiv:1611.03668 [gr-qc]].

\bibitem{Donnelly:2016auv} 
W.~Donnelly and L.~Freidel,
``Local subsystems in gauge theory and gravity,''
JHEP {\bf 1609}, 102 (2016)
doi:10.1007/JHEP09(2016)102
[arXiv:1601.04744 [hep-th]].



\bibitem{Bondi:1962px} 
  H.~Bondi, M.~G.~J.~van der Burg and A.~W.~K.~Metzner,
  ``Gravitational waves in general relativity. 7. Waves from axisymmetric isolated systems,''
  Proc.\ Roy.\ Soc.\ Lond.\ A {\bf 269}, 21 (1962).
  doi:10.1098/rspa.1962.0161


\bibitem{Sachs:1962wk} 
  R.~K.~Sachs,
  ``Gravitational waves in general relativity. 8. Waves in asymptotically flat space-times,''
  Proc.\ Roy.\ Soc.\ Lond.\ A {\bf 270}, 103 (1962).
  doi:10.1098/rspa.1962.0206


\bibitem{Arnowitt:1962hi} 
R.~L.~Arnowitt, S.~Deser and C.~W.~Misner,
``The Dynamics of general relativity,''
Gen.\ Rel.\ Grav.\  {\bf 40}, 1997 (2008)
[gr-qc/0405109].






\bibitem{Poisson} 
E. Poisson, ``A relativists' toolkit," Cambridge (2004)



\bibitem{Higuchi:1991tk} 
  A.~Higuchi,
  ``Quantum linearization instabilities of de Sitter space-time. 1,''
  Class.\ Quant.\ Grav.\  {\bf 8}, 1961 (1991).
  doi:10.1088/0264-9381/8/11/009


\bibitem{Saunders} 
J. Saunders, ``The Geometry of jet bundles," Cambridge (1989)



\bibitem{Park:2014qoa} 
I.~Y.~Park,
``Quantization of gravity through hypersurface foliation,''
arXiv:1406.0753 [gr-qc].


\bibitem{Park:2015qxa} 
I.~Y.~Park,
``Foliation, jet bundle and quantization of Einstein gravity,''
Front.\ in Phys.\  {\bf 4}, 25 (2016)
doi:10.3389/fphy.2016.00025
[arXiv:1503.02015 [hep-th]].




\bibitem{York:1972sj} 
J.~W.~York, Jr.,
``Role of conformal three geometry in the dynamics of gravitation,''
Phys.\ Rev.\ Lett.\  {\bf 28}, 1082 (1972).



\bibitem{Moncrief:1989dx} 
V.~Moncrief,
``Reduction of the Einstein equations in (2+1)-dimensions to a Hamiltonian system over Teichmuller space,''
J.\ Math.\ Phys.\  {\bf 30}, 2907 (1989).


\bibitem{Fischer:1996qg} 
A.~E.~Fischer and V.~Moncrief,
``Hamiltonian reduction of Einstein's equations of general relativity,''
Nucl.\ Phys.\ Proc.\ Suppl.\  {\bf 57}, 142 (1997).


\bibitem{Marsden}
J. E. Marsden, G. Misiolek, J.-P. Ortega, M. Perlmutter, and T. S. Ratiu, ``Hamiltonian Reduction by Stages," Spinger (2007)




\bibitem{Gay-Balmaz:2014ena} 
F.~Gay-Balmaz and T.~S.~Ratiu,
``A new Lagrangian dynamic reduction in field theory,''
Annales Inst.\ Fourier {\bf 16}, 1125 (2010)
[arXiv:1407.0263 [math-ph]].





\bibitem{deBoer:2003vf} 
  J.~de Boer and S.~N.~Solodukhin,
  ``A Holographic reduction of Minkowski space-time,''
  Nucl.\ Phys.\ B {\bf 665}, 545 (2003)
  doi:10.1016/S0550-3213(03)00494-2
  [hep-th/0303006].

\bibitem{Gomes:2010fh} 
  H.~Gomes, S.~Gryb and T.~Koslowski,
  ``Einstein gravity as a 3D conformally invariant theory,''
  Class.\ Quant.\ Grav.\  {\bf 28}, 045005 (2011)
  doi:10.1088/0264-9381/28/4/045005
  [arXiv:1010.2481 [gr-qc]].





\bibitem{Gourgoulhon} 
E. Gourgoulhon, ``3+1 Formalism in general relativity," Springer (2012)





\bibitem{Landau} 
L D Landau and E.M. Lifshitz, ``The Classical Theory of Fields," 4th ed, Butterworth Heinemann (1975)



\bibitem{Shapiro} 
T. W. Baumgarte and S. L. Shapiro, ``Numerical relativity," Cambridge (2010)



\bibitem{Isham} 
C. J. Isham, ``Modern differential geometry for physicists," 2nd ed, World Scientific (1999)

\bibitem{Eichhorn:2013ug} 
  A.~Eichhorn,
  ``Faddeev-Popov ghosts in quantum gravity beyond perturbation theory,''
  Phys.\ Rev.\ D {\bf 87}, no. 12, 124016 (2013)
  doi:10.1103/PhysRevD.87.124016
  [arXiv:1301.0632 [hep-th]].

\bibitem{Aliev:2004ds} 
  A.~N.~Aliev and A.~E.~Gumrukcuoglu,
  ``Gravitational field equations on and off a 3-brane world,''
  Class.\ Quant.\ Grav.\  {\bf 21}, 5081 (2004)
  doi:10.1088/0264-9381/21/22/005
  [hep-th/0407095].




\bibitem{Park:2013iqa} 
  I.~Y.~Park,
  ``ADM reduction of Einstein action and black hole entropy,''
  Fortsch.\ Phys.\  {\bf 62}, 950 (2014)
  doi:10.1002/prop.201400056
  [arXiv:1304.0014 [hep-th]].

\bibitem{Park:2013vpa} 
  I.~Y.~Park,
  ``Dimensional reduction to hypersurface of foliation,''
  Fortsch.\ Phys.\  {\bf 62}, 966 (2014)
  doi:10.1002/prop.201400068
  [arXiv:1310.2507 [hep-th]].




\bibitem{Park:2013bma} 
  I.~Y.~Park,
  ``Reduction of BTZ spacetime to hypersurfaces of foliation,''
  JHEP {\bf 1401}, 102 (2014)
  doi:10.1007/JHEP01(2014)102
  [arXiv:1311.4619 [hep-th]].



\bibitem{VanNieuwenhuizen:1981ae} 
  P.~Van Nieuwenhuizen,
  ``Supergravity,''
  Phys.\ Rept.\  {\bf 68}, 189 (1981).
  doi:10.1016/0370-1573(81)90157-5

\bibitem{Duff:1986hr} 
  M.~J.~Duff, B.~E.~W.~Nilsson and C.~N.~Pope,
  ``Kaluza-Klein Supergravity,''
  Phys.\ Rept.\  {\bf 130}, 1 (1986).
  doi:10.1016/0370-1573(86)90163-8







\bibitem{Sato:2002kv} 
  M.~Sato and A.~Tsuchiya,
  ``Born-Infeld action from supergravity,''
  Prog.\ Theor.\ Phys.\  {\bf 109}, 687 (2003)
  doi:10.1143/PTP.109.687
  [hep-th/0211074].

\bibitem{Hatefi:2012bp} 
  E.~Hatefi, A.~J.~Nurmagambetov and I.~Y.~Park,
  ``ADM reduction of IIB on $\mathcal{H}^{p,q}$ and dS braneworld,''
  JHEP {\bf 1304}, 170 (2013)
  doi:10.1007/JHEP04(2013)170
  [arXiv:1210.3825 [hep-th]].  


\bibitem{Niarchos:2015moa} 
V.~Niarchos,
``Open/closed string duality and relativistic fluids,''
Phys.\ Rev.\ D {\bf 94}, no. 2, 026009 (2016)
doi:10.1103/PhysRevD.94.026009
[arXiv:1510.03438 [hep-th]].


\bibitem{Grignani:2016bpq} 
G.~Grignani, T.~Harmark, A.~Marini and M.~Orselli,
``The Born-Infeld/Gravity Correspondence,''
arXiv:1602.01640 [hep-th].

\bibitem{Maxfield:2016vpw} 
  T.~Maxfield and S.~Sethi,
  ``DBI from Gravity,''
  JHEP {\bf 1702}, 108 (2017)
  doi:10.1007/JHEP02(2017)108
  [arXiv:1612.00427 [hep-th]].



\bibitem{Park:2008sg} 
  I.~Y.~Park,
  ``One loop scattering on D-branes,''
  Eur.\ Phys.\ J.\ C {\bf 62}, 783 (2009)
  doi:10.1140/epjc/s10052-009-1065-4
  [arXiv:0801.0218 [hep-th]].



\bibitem{Park:2015ota} 
  I.~Y.~Park,
  ``4D covariance of holographic quantization of Einstein gravity,''
  arXiv:1506.08383 [hep-th].


\bibitem{Geroch:1972up} 
  R.~P.~Geroch,
  ``Structure of the gravitational field at spatial infinity,''
  J.\ Math.\ Phys.\  {\bf 13}, 956 (1972).
  doi:10.1063/1.1666094


\bibitem{Ashtekar:1978zz} 
  A.~Ashtekar and R.~O.~Hansen,
  ``A unified treatment of null and spatial infinity in general relativity. I - Universal structure, asymptotic symmetries, and conserved quantities at spatial infinity,''
  J.\ Math.\ Phys.\  {\bf 19}, 1542 (1978).
  doi:10.1063/1.523863


\bibitem{Arcioni:2003td} 
  G.~Arcioni and C.~Dappiaggi,
  ``Holography in asymptotically flat space-times and the BMS group,''
  Class.\ Quant.\ Grav.\  {\bf 21}, 5655 (2004)
  doi:10.1088/0264-9381/21/23/022
  [hep-th/0312186].



\bibitem{Carney:2017jut} 
  D.~Carney, L.~Chaurette, D.~Neuenfeld and G.~W.~Semenoff,
  ``Infrared quantum information,''
  arXiv:1706.03782 [hep-th].


\bibitem{Strominger:2017aeh} 
  A.~Strominger,
  ``Black Hole Information Revisited,''
  arXiv:1706.07143 [hep-th];
  ``Lectures on the Infrared Structure of Gravity and Gauge Theory,''
  arXiv:1703.05448 [hep-th].



\bibitem{Bousso:2017dny} 
  R.~Bousso and M.~Porrati,
  ``Soft Hair as a Soft Wig,''
  arXiv:1706.00436 [hep-th].







%
%
%
%

%
%
%
%
%

































%
%
%
%
%













\bibitem{Callan:1977pt} 
  C.~G.~Callan, Jr. and S.~R.~Coleman,
  ``The Fate of the False Vacuum. 2. First Quantum Corrections,''
  Phys.\ Rev.\ D {\bf 16}, 1762 (1977).
  doi:10.1103/PhysRevD.16.1762





\bibitem{EWeinberg}

E. Weinberg, ``Classical solutions in quantum field theory" Cambridge university press (2012)





\bibitem{Park:2017dib} 
  I.~Y.~Park,
  ``Quantum-corrected geometry of horizon vicinity,''
  arXiv:1704.04685 [hep-th].




\bibitem{Park:2016vam} 
  F.~James and I.~Y.~Park,
  ``Quantum gravitational effects on boundary,''
  arXiv:1610.06464 [hep-th], to appear in TMPh.















%
%
%
%










%
%
%
%
%
%
%














































\end{thebibliography}
\end{document}